\documentstyle[12pt]{article}

\begin{document}
\title{Method of Covariant Calculation of the Amplitudes of
Processes with the Polarized Dirac Particles\footnote
{
Presented at the Joint International Workshop: VIII Workshop on
High Energy Physics and Quantum Field Theory \& III Workshop on
Physics at VLEPP, Zvenigorod, Russia, September 15--21, 1993
} }

\author {Alexander L. Bondarev
\and \it National Scientific and Educational Center of Particle and
\and \it High Energy Physics attached to Belarusian State University
\and \it M.Bogdanovich str.,153, Minsk 220040, Republic of Belarus
\and \rm e-mail: bondarev@hep.by}

\setlength{\textheight}{218mm}
\setlength{\textwidth}{150mm}

\maketitle

\begin{abstract}
    General  scheme  for  covariant  calculation  of  the  amplitudes of
processes with the  polarized Dirac particles  is considered.   It is so
concretized that the obtained expressions can be used for calculation of
the  amplitudes  of  processes   with  interfering  diagrams.     As  an
illustration  the  expressions  for  the  amplitudes  of  processes with
massless particles are presented.
\end{abstract}

\maketitle

\section {Introduction}

    The serious  difficulties take  place when  we calculate the
cross sections of processes for high order diagrams  (especially
when we take into  account polarizations of particles,  involved
in  reaction),  because  it  is  necessary to evaluate traces of
products of the great number of Dirac $\gamma $-matrices.  In this
case  the  problem  often  arises  with  obtaining of analytical
expressions for the different physical quantities because of their
cumbersomeness .

    One of the  ways to avoid  this problem is  to calculate the
amplitudes  of  the  processes  directly.    In  particular,   the
expressions, obtained  by multiplication  of $\gamma  $-matrices
and bispinors which are written in components in concrete  frames,
are given in \cite  {r1}.  Authors of  some later articles had  to
use such  a method  on account  of calculating  difficulties (see,
e.g.    \cite  {r2}).    The  obvious  defects  of this method are
complicated calculations, bulky and noncovariant form of  obtained
results.

    Different authors made attempts of the covariant calculation
of  amplitudes   (see  \cite   {r3} -- \cite   {r8}).    However,
expressions   obtained   in   their   papers  are  unsuitable  for
calculations with interfering diagrams.  The scheme which is the
extension of the results of these authors is considered below.  It
is concretized to avoid problems with interference.

\section {General scheme of covariant calculation of amplitudes}

    There is even number $(2N)$ of fermions in initial and final
state  for  any  reaction  with  Dirac particles.  Therefore every
diagram contains $N$ nonclosed fermion lines.  The expression
$$
\displaystyle
 M_{if} = \bar{u}_f Q u_i
$$
    corresponds to every line in the amplitude of process, where
$\; u_i, \;  u_f \;$  are  Dirac  bispinors  for free particles.
(For definiteness, we anticipate that fermions are particles.
However the  obtained results  are true  in case if both fermions
are antiparticles  or one fermion is  particle and another one is
antiparticle.)
$$
\bar{u} = u^{+} {\gamma}^0 \;\;.
$$
    $Q$ is matrix operator which characterizes interaction.   It
is  expressed  as  linear  combination  of  the  products of Dirac
$\gamma $ -matrices  (or of its  contractions with 4-vectors)  and
can have any number of free Lorentz indexes.

For calculating $M_{if}$ we use the following scheme:
\begin {equation}
\begin {array}{l} \displaystyle
M_{if} = \bar{u}_f Q u_i = ( \bar{u}_f Q u_i ) \cdot
       { \bar{u}_i Z u_f \over \bar{u}_i Z u_f }
= { Tr ( Q u_i \bar{u}_i Z u_f \bar{u}_f ) \over
     \bar{u}_i Z u_f }
            \\[0.5cm] \displaystyle
\simeq { Tr ( Q u_i \bar{u}_i Z u_f \bar{u}_f ) \over
         | \bar{u}_i Z u_f | }
= { Tr ( Q u_i \bar{u}_i Z u_f \bar{u}_f ) \over
  [ Tr ( \bar{Z} u_i \bar{u}_i Z u_f \bar{u}_f ) ]^{1/2} }
   = {\cal M}_{if}
\end {array}
\label{e1}
\end {equation}
where $Z$ is an arbitrary $4 \times 4$-matrix ,
$$
\bar{Z} = {\gamma}^0 Z^{+} {\gamma}^0
$$
(the symbol $\simeq$  stands for "an equality to within a phase
factor sign").

The projection operators are substituted for $u \bar{u}$ in
(\ref{e1}). For particle with mass $m$:
\begin {equation}
\displaystyle
u(p,n) \bar{u}(p,n) = { 1 \over 4m } ( \hat{p} + m )
                ( 1 + {\gamma}_5 \hat{n} ) = {\cal P}
\label{e2}
\end {equation}
where
$
\displaystyle
\hat{p} = {\gamma}_{\mu} p^{\mu}, \;\; p^2 = m^2, \;\;
n^2 = -1, \;\; pn = 0
\bar{u} u = 1, \;\;
{\gamma}_5 = {\it i} {\gamma}^0 {\gamma}^1 {\gamma}^2 {\gamma}^3
\;\;.
$
[We use the  metric
$ \;\;
\displaystyle
a^{\mu} = ( a_0, \vec{a} ),
\;\;
a_{\mu} = ( a_0, -\vec{a} ), \;\;
ab = a_{\mu} b^{\mu} = a_0 b_0 - \vec{a} \vec{b} \; .]
$

For massless particle the projection operator is the following:
\begin {equation}
\displaystyle
u_{\pm}(q) \bar{u}_{\pm}(q)
= { 1 \over 2 }( 1 \pm {\gamma}_5 ) \hat{q}
= {\cal P}_{\pm }
\label{e3}
\end {equation}
where
$
\displaystyle
q^2 = 0, \;\; \bar{u}_{\pm} {\gamma}_{\mu} u_{\pm} = 2 q_{\mu}
$
(signs $\pm $ correspond to helicity of particle).

    Notice that
\begin {equation}
\begin {array}{c} \displaystyle
( {\cal M}_{if} )^{*}
= { [ ( \bar{u}_f Q u_i ) ( \bar{u}_i Z u_f ) ]^{*} \over
    [ Tr ( \bar{Z} u_i \bar{u}_i Z u_f \bar{u}_f ) ]^{1/2} }
= { (\bar{u}_f \bar{Z} u_i ) ( \bar{u}_i \bar{Q} u_f ) \over
    [ Tr ( \bar{Z} u_i \bar{u}_i Z u_f \bar{u}_f ) ]^{1/2} }
           \\[0.5cm] \displaystyle
= { Tr ( \bar{Z} u_i \bar{u}_i \bar{Q} u_f \bar{u}_f ) \over
    [ Tr ( \bar{Z} u_i \bar{u}_i Z u_f \bar{u}_f ) ]^{1/2} } \;\;.
\end {array}
\label{e4}
\end {equation}

    In the articles \cite{r3}, \cite{r4} one chooses
$$
\displaystyle
Z=1 \;\;.
$$
The  results  of  article  \cite{r5}  come  to the same approach for
the processes with massless particles.

    In \cite{r4} one proposes
$$
\displaystyle
Z = {\gamma}_5 \;\;
$$
too.  The results of article \cite{r6} come to the same choice.

    The results of article \cite{r7} correspond to the choice
$$
\displaystyle
Z = 1 + {\gamma}^0 \;\;.
$$

   The results of \cite{r8} come to
$$
\displaystyle
Z = m + \hat{r}
$$
(where $r$ is arbitrary 4-momentum, such as $r^2 = m^2$). In this
paper for 4-vectors, which determine axes of spin projections,  one
uses
$$
\displaystyle
n_i = { m^2 p_f - (p_i p_f) p_i \over
       m [ (p_i p_f)^2 - m^4 ]^{1/2} } \; , \;\;
n_f = - { m^2 p_i -(p_i p_f) p_f \over
        m [ (p_i p_f)^2 - m^4 ]^{1/2} } \; .
$$

    However all expressions for the amplitudes [as it follows from
(\ref{e1})] are known to within a phase factor.  Really
$$
\displaystyle
{\cal M}_{if} = M_{if} \cdot {\bar{u}_i Z u_f \over
                            | \bar{u}_i Z u_f | } \;\; .
$$
It is obvious that this circumstance creates no problems, when
we calculate amplitude for one  diagram.  However, in the  general
case,  the  presence  of  unknown  phase  factor does not enable
formula (\ref{e1}) to be used for calculation of amplitudes of the
processes which proceed  in a few  channels since the  expressions
for  amplitudes  which  corresponding  to different channels are
multiplied by different phase factors.

\section {Calculation of the amplitudes of processes with interfering
diagrams}

    Let us  consider in the  general form  the process,  which
proceeds in two different channels (see Fig.\ref{Fg1}):
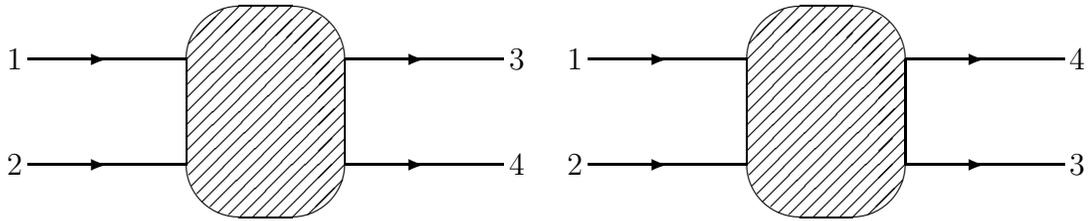
\begin{figure}[ht]
\begin{tabular}{cc}
\begin{picture}(200,100)
\put(100,50){\oval(60,80)}
\put(71,76){\line(1,1){13.}}
\put(70,70){\line(1,1){20.}}
\put(70,65){\line(1,1){25.}}
\put(70,60){\line(1,1){30.}}
\put(70,55){\line(1,1){35.}}
\put(70,50){\line(1,1){40.}}
\put(70,45){\line(1,1){44.}}
\put(70,40){\line(1,1){48.}}
\put(70,35){\line(1,1){51.}}
\put(70,30){\line(1,1){54.}}
\put(71,26){\line(1,1){55.}}
\put(72,22){\line(1,1){56.}}
\put(74,19){\line(1,1){55.}}
\put(76,16){\line(1,1){54.3}}
\put(79,14){\line(1,1){51.}}
\put(82,12){\line(1,1){48.}}
\put(86,11){\line(1,1){44.}}
\put(90,10){\line(1,1){40.}}
\put(95,10){\line(1,1){35.}}
\put(100,10){\line(1,1){30.}}
\put(105,10){\line(1,1){25.}}
\put(110,10){\line(1,1){20.}}
\put(116,11){\line(1,1){13.}}
\put(05,30){\makebox(0,0){$ 2 $}}
\put(05,70){\makebox(0,0){$ 1 $}}
\put(195,30){\makebox(0,0){$ 4 $}}
\put(195,70){\makebox(0,0){$ 3 $}}
\thicklines
\put(10,30){\line(2,0){60.}}
\thicklines
\put(10,70){\line(2,0){60.}}
\thicklines
\put(130,30){\line(2,0){60.}}
\thicklines
\put(130,70){\line(2,0){60.}}
\put(40,30){\vector(1,0){0.}}
\put(40,70){\vector(1,0){0.}}
\put(160,30){\vector(1,0){0.}}
\put(160,70){\vector(1,0){0.}}
\end{picture}
&
\begin{picture}(200,100)
\put(100,50){\oval(60,80)}
\put(71,76){\line(1,1){13.}}
\put(70,70){\line(1,1){20.}}
\put(70,65){\line(1,1){25.}}
\put(70,60){\line(1,1){30.}}
\put(70,55){\line(1,1){35.}}
\put(70,50){\line(1,1){40.}}
\put(70,45){\line(1,1){44.}}
\put(70,40){\line(1,1){48.}}
\put(70,35){\line(1,1){51.}}
\put(70,30){\line(1,1){54.}}
\put(71,26){\line(1,1){55.}}
\put(72,22){\line(1,1){56.}}
\put(74,19){\line(1,1){55.}}
\put(76,16){\line(1,1){54.3}}
\put(79,14){\line(1,1){51.}}
\put(82,12){\line(1,1){48.}}
\put(86,11){\line(1,1){44.}}
\put(90,10){\line(1,1){40.}}
\put(95,10){\line(1,1){35.}}
\put(100,10){\line(1,1){30.}}
\put(105,10){\line(1,1){25.}}
\put(110,10){\line(1,1){20.}}
\put(116,11){\line(1,1){13.}}
\put(05,30){\makebox(0,0){$ 2 $}}
\put(05,70){\makebox(0,0){$ 1 $}}
\put(195,30){\makebox(0,0){$ 3 $}}
\put(195,70){\makebox(0,0){$ 4 $}}
\thicklines
\put(10,30){\line(2,0){60.}}
\thicklines
\put(10,70){\line(2,0){60.}}
\thicklines
\put(130,30){\line(2,0){60.}}
\thicklines
\put(130,70){\line(2,0){60.}}
\put(40,30){\vector(1,0){0.}}
\put(40,70){\vector(1,0){0.}}
\put(160,30){\vector(1,0){0.}}
\put(160,70){\vector(1,0){0.}}
\end{picture}
\end{tabular}
\caption{ The diagrams of the process, which proceeds in 2 different
channels in general form.}
\label{Fg1}
\end{figure}

The expression
$$
\displaystyle
M = ( \bar{u}_3 Q u_1 ) \cdot ( \bar{u}_4 R u_2 )
= M_{13} \cdot M_{24}
$$
corresponds to the first diagram. The expression
$$
\displaystyle
 M' = ( \bar{u}_4 S u_1 ) \cdot ( \bar{u}_3 T u_2 )
            = M_{14} \cdot M_{23}
$$
corresponds to the second one, where $Q$, $R$, $S$, $T$
are the arbitrary matrix operators characterizing interaction.

    There are three possibilities  for the correct calculation
of the amplitude of process (see \cite{r4}):
\begin{enumerate}
%1
\item
     It  is possible  to make  Fierz arrangement  for the second
diagram:   $3 \leftrightarrow  4$.   But this  method requires a
large volume of additional calculations.
%2
\item
   It  is possible  to multiply  both amplitudes  by the  same phase
factors, for example by
$$
\displaystyle
{ \bar{u}_1 Z u_3 \over | \bar{u}_1 Z u_3 | } \cdot
{ \bar{u}_2 X u_4 \over | \bar{u}_2 X u_4 | } \;\;\; .
$$
We  have expression for the first diagram in this case
$$
\displaystyle
   { Tr ( Q u_1 \bar{u}_1 Z u_3 \bar{u}_3 ) \over
   [ Tr ( \bar{Z} u_1 \bar{u}_1 Z u_3 \bar{u}_3 ) ]^{1/2} } \cdot
   { Tr ( R u_2 \bar{u}_2 X u_4 \bar{u}_4 ) \over
   [ Tr ( \bar{X} u_2 \bar{u}_2 X u_4 \bar{u}_4 ) ]^{1/2} }
$$
and that for the second diagram
$$
\displaystyle
{ Tr ( S u_1 \bar{u}_1 Z u_3 \bar{u}_3 T u_2 \bar{u}_2
     X u_4 \bar{u}_4 ) \over
[ Tr ( \bar{Z} u_1 \bar{u}_1 Z u_3 \bar{u}_3 ) ]^{1/2} \cdot
 [Tr ( \bar{X} u_2 \bar{u}_2 X u_4 \bar{u}_4 ) ]^{1/2} } \;\; .
$$
    It is obvious that calculation of the amplitude for the second
diagram is very complicated.  Besides, it is very inconvenient  to
use expressions  with the  different structure  for calculation of
amplitudes for the different diagrams.
%3
\item
    The third possibility is to calculate the relative  phase
for  the  first  and  second  diagrams  and  to use expression the
obtained for the phase correction of one of two amplitudes.
\end{enumerate}

We will use the third possibility. Interference term has the form
$$
\displaystyle
M \cdot (M')^{*}
= M_{13} \cdot M_{24} \cdot ( M_{14} )^{*} \cdot ( M _{23} )^{*}
= ( \bar{u}_3 Q u_1 ) ( \bar{u}_4 R u_2 )
  ( \bar{u}_1 \bar{S} u_4 ) ( \bar{u}_2 \bar{T} u_3 ) \;\; .
$$

Let us multiply the interference term by
$$
\displaystyle
   { ( \bar{u}_1 Z u_3 ) ( \bar{u}_2 Y u_4 )
     ( \bar{u}_4 \bar{X} u_1 ) ( \bar{u}_3 \bar{V} u_2 ) \over
     ( \bar{u}_1 Z u_3 ) ( \bar{u}_2 Y u_4 )
     ( \bar{u}_4 \bar{X} u_1 ) ( \bar{u}_3 \bar{V} u_2 ) }
\cdot { ( \bar{u}_3 \bar{Z} u_1 ) ( \bar{u}_1 X u_4 )
        ( \bar{u}_4 \bar{Y} u_2 ) ( \bar{u}_2 V u_3 ) \over
        ( \bar{u}_3 \bar{Z} u_1 ) ( \bar{u}_1 X u_4 )
        ( \bar{u}_4 \bar{Y} u_2 ) ( \bar{u}_2 V u_3 ) } \equiv 1
$$
where $X$, $Y$, $Z$, $V$ are  as  yet  arbitrary matrix operators.

We have as the result in this case:
$$
\begin {array}{l} \displaystyle
M \cdot (M')^{*}
                        \\[0.5cm] \displaystyle
\equiv { Tr ( Q u_1 \bar{u}_1 Z u_3 \bar{u}_3 ) \over
         Tr ( \bar{Z} u_1 \bar{u}_1 Z u_3 \bar{u}_3 ) } \cdot
  { Tr ( R u_2 \bar{u}_2 Y u_4 \bar{u}_4) \over
    Tr ( \bar{Y} u_2 \bar{u}_2 Y u_4 \bar{u}_4 ) } \cdot
{ Tr ( \bar{X} u_1 \bar{u}_1 \bar{S} u_4 \bar{u}_4 ) \over
         Tr ( \bar{X} u_1 \bar{u}_1 X u_4 \bar{u}_4 ) } \cdot
   { Tr ( \bar{V} u_2 \bar{u}_2 \bar{T} u_3 \bar{u}_3 ) \over
     Tr ( \bar{V} u_2 \bar{u}_2 V u_3 \bar{u}_3 ) }
                        \\[0.5cm] \displaystyle
\times Tr ( \bar{Z} u_1 \bar{u}_1 X u_4 \bar{u}_4
          \bar{Y} u_2 \bar{u}_2 V u_3 \bar{u}_3 )
                        \\[0.5cm] \displaystyle
= { Tr ( Q u_1 \bar{u}_1 Z u_3 \bar{u}_3 ) \over
  [ Tr ( \bar{Z} u_1 \bar{u}_1 Z u_3 \bar{u}_3 ) ]^{1/2} }
 \cdot { Tr(R u_2 \bar{u}_2 Y u_4 \bar{u}_4 ) \over
       [ Tr ( \bar{Y} u_2 \bar{u}_2 Y u_4 \bar{u}_4 ) ]^{1/2} }
                        \\[0.5cm] \displaystyle
\times { Tr ( \bar{X} u_1 \bar{u}_1 \bar{S} u_4 \bar{u}_4 ) \over
       [ Tr ( \bar{X} u_1 \bar{u}_1 X u_4 \bar{u}_4 ) ]^{1/2} }
 \cdot { Tr ( \bar{V} u_2 \bar{u}_2 \bar{T} u_3 \bar{u}_3 )
 \over [ Tr ( \bar{V} u_2 \bar{u}_2 V u_3 \bar{u}_3 ) ]^{1/2} }
                        \\[0.5cm] \displaystyle
\times { Tr ( \bar{Z} u_1 \bar{u}_1 X u_4 \bar{u}_4
           \bar{Y} u_2 \bar{u}_2 V u_3 \bar{u}_3 ) \over
 [ Tr ( \bar{Z} u_1 \bar{u}_1 Z u_3 \bar{u}_3 )
   Tr ( \bar{X} u_1 \bar{u}_1 X u_4 \bar{u}_4 )
   Tr ( \bar{Y} u_2 \bar{u}_2 Y u_4 \bar{u}_4 )
   Tr ( \bar{V} u_2 \bar{u}_2 V u_3 \bar{u}_3 ) ]^{1/2} }
                        \\[0.5cm] \displaystyle
= {\cal M}_{13} \cdot {\cal M}_{24} \cdot ( {\cal M}_{14} )^{*}
    \cdot ( {\cal M}_{23} )^{*} \cdot K
\end {array}
$$
where
$
\displaystyle
{\cal M}_{13}, {\cal M}_{24}, ( {\cal M}_{14} )^{*},
( {\cal M}_{23} )^{*}
$
are  given by  expressions  analogous to
(\ref{e1}), (\ref{e4}); the coefficient $K$ is given by formula
\begin {equation}
\displaystyle
K = { Tr ( \bar{Z} u_1 \bar{u}_1 X u_4 \bar{u}_4
       \bar{Y} u_2 \bar{u}_2 V u_3 \bar{u}_3 ) \over
    [ Tr ( \bar{Z} u_1 \bar{u}_1 Z u_3 \bar{u}_3 )
      Tr ( \bar{X} u_1 \bar{u}_1 X u_4 \bar{u}_4 )
      Tr ( \bar{Y} u_2 \bar{u}_2 Y u_4 \bar{u}_4 )
      Tr( \bar{V} u_2 \bar{u}_2 V u_3 \bar{u}_3 ) ]^{1/2} }.
\label{e5}
\end {equation}
It is obvious that
$$
\displaystyle
|K| = 1.
$$
    Thus  we  have to calculate the amplitude  of  process with
interfering diagrams in the form
\begin {equation}
\displaystyle
{\cal M} + {\cal M}'
= K \cdot {\cal M}_{13} \cdot {\cal M}_{24}
        + {\cal M}_{14} \cdot {\cal M}_{23} \;\; .
\label{e6}
\end {equation}

Let us require for the maximum simplicity of calculations
$$
 K \equiv 1 .
$$
This requirement is satisfied if we choose  \\
$Z = X = Y = V = {\cal P} \; $ [see  (\ref{e2})]
$\;\;\;$ or $\;\;\;$
$Z = X = Y = V = {\cal P}_{\pm} \; $ [see  (\ref{e3})], \\
since the projection operators have the following properties
\begin {equation}
\displaystyle
\bar{\cal P} = {\cal P} \;\; , \;\;\;\;\;
{\cal P} A {\cal P} = Tr [ {\cal P} A ] \cdot {\cal P} \;\; ,
\label{e7}
\end {equation}
\begin {equation}
\displaystyle
\bar{\cal P}_{\pm} = {\cal P}_{\pm} \;\; , \;\;\;\;\;
{\cal P}_{\pm} A {\cal P}_{\pm}
= Tr [ {\cal P}_{\pm} A ] \cdot {\cal P}_{\pm}
\;\;\; .
\label{e8}
\end {equation}
Really
$$
\begin {array}{l} \displaystyle
\bar{\cal P} = {\gamma}^0 {\cal P}^{+} {\gamma}^0
= {\gamma}^0 ( u \bar{u} )^{+} {\gamma}^0
= {\gamma}^0 ( u u^{+} {\gamma}^0 )^{+} {\gamma}^0
= {\gamma}^0 [ ( {\gamma}^0 )^{+} ( u^{+} )^{+} u^{+} ] {\gamma}^0
                   \\[0.5cm] \displaystyle
= {\gamma}^0 [ {\gamma}^0 u u^{+} ] {\gamma}^0
= u u^{+} {\gamma}^0 = u \bar{u} = {\cal P} \;\; ,
\end {array}
$$
$$
\begin {array}{l} \displaystyle
{\cal P} A {\cal P} = (u)_{\alpha} ( \bar{u} )_{\beta}
(A)^{\beta \rho} (u)_{\rho} ( \bar{u} )_{\delta}
= [ ( \bar{u} )_{\beta } (A)^{\beta \rho} (u)_{\rho} ]
(u)_{\alpha } ( \bar{u} )_{\delta}
                   \\[0.5cm] \displaystyle
= [ (u)_{\rho} ( \bar{u} )_{\beta} (A)^{\beta \rho} ]
    (u)_{\alpha} ( \bar{u} )_{\delta}
= Tr [ {\cal P} A ] \cdot {\cal P} \;\; .
\end {array}
$$

Therefore we can calculate the amplitude
\begin {equation}
\begin {array}{r} \displaystyle
{\cal M} + {\cal M}'
= { Tr ( Q {\cal P}_1 {\cal P} {\cal P}_3 ) \cdot
    Tr ( R {\cal P}_2 {\cal P} {\cal P}_4) \over
  [ Tr ( {\cal P} {\cal P}_1 ) Tr ( {\cal P} {\cal P}_2 )
  Tr ( {\cal P} {\cal P}_3 ) Tr ( {\cal P} {\cal P}_4 ) ]^{1/2} }
                   \\[0.5cm] \displaystyle
  + { Tr ( S {\cal P}_1 {\cal P} {\cal P}_4 ) \cdot
      Tr ( T {\cal P}_2 {\cal P} {\cal P}_3 ) \over
    [ Tr ( {\cal P} {\cal P}_1 ) Tr ( {\cal P} {\cal P}_2 )
      Tr ( {\cal P} {\cal P}_3 )
      Tr ( {\cal P} {\cal P}_4 ) ]^{1/2} } \;\; .
\end {array}
\label{e9}
\end {equation}
    This   expression   enables   to   calculate   the amplitude
numerically.  Complex numbers being obtained  under  calculation
are used for calculation of the process cross section.

    Notice that the calculation of amplitude for the alone diagram
is easer than the calculation of squared amplitude, if  operators,
characterizing  inter\-action,  contain  the  product  of  greater
number of $\gamma $-matrices, than the projection operator.

    Really, when the number of $\gamma $-matrices in operator  $Q$
increases by $I$ [see  (\ref{e1})], their number in  the numerator
of (\ref{e1}) increases only by $I$ (denominator does not change),
but  in  construction  of
$  Tr  (  Q  u_i  \bar{u}_i \bar{Q} u_f
\bar{u}_f  )
$ ,
which  appears,  when  we   calculate  the squared matrix element,
the number of $\gamma $-matrices increases by $2I$.  We take  into
account that trace of product of $2J$ $\gamma $-matrices  contains
$
1\cdot 3\cdot  5\cdot \ldots  \cdot (2J-1)
$
terms  and  we  obtain  that the more complicated is a process the
bigger are the advantage  given under calculation of  this process
by the method of direct amplitudes calculation.

    However, for processes with the interfering diagrams  method
of amplitudes calculation is easier in any case, because we need
not calculate the interference terms.

    As  it  was  mentioned  before,   it  is  simple  to  make   a
generalization of this method for the reactions with participation
of  antiparticles.    It  is  sufficient  for it to substitute the
projection operators  of antiparticles  in place  of operators  of
particles.    Let,  for  example,  we  are  interested  in   value
$
\displaystyle
\bar{v}_f  Q  u_i
$ ,
where   $v_f$  is  bispinor  for  a   free
antiparticle.  Then
\begin {equation}
\displaystyle
\bar{v}_f Q u_i = { Tr ( Q u_i \bar{u}_i Z v_f \bar{v}_f )
\over [ Tr ( \bar{Z} u_i \bar{u}_i Z v_f \bar{v}_f ) ]^{1/2} }
\label{e10}
\end {equation}
where
$$
\displaystyle
v(p,n) \bar{v}(p,n)
= { 1 \over 4m } ( -m + \hat{p} ) ( 1 + {\gamma}_5 \hat{n} )
$$
for massive antiparticle, or
$$
\displaystyle
v_{\pm}(q) \bar{v}_{\pm}(q)
= { 1 \over 2 }( 1 \mp {\gamma}_5) \hat{q}
$$
for massless antiparticle.
As always, we use  (\ref{e2}) or (\ref{e3}) instead of $Z$ .

    Notice that in (\ref{e10}) and in the further  consideration
we shall  use equality  sign instead  of symbol  $\simeq $ , since
there exists not any trouble with the phase factors already.

    It  is  significant  that  the  methods  of calculation of the
amplitudes with $Z = 1, {\gamma}_5, 1 + {\gamma}^0, m + \hat{p} $
for the processes with  interfering diagrams require to  calculate
the expression not only for amplitudes of separate diagrams, but
also for  coefficient $K$ by  the formula  (\ref{e5}).   It is
necessary to use formula (\ref{e6}) in this case.

\section {Application of method to calculations  of the amplitudes
of processes with massless Dirac particles}

    Formulas for calculation of amplitudes of processes with the
massless Dirac particles  are very simple.   In this  case formula
(\ref{e1})   takes   the   following   form
\begin  {equation}
\displaystyle
\bar{u}_{\pm} (p_3) Q u_{\pm} (p_1)
= { Tr [ Q \hat{p}_1 \hat{q} \hat{p}_3 ( 1 \mp {\gamma}_5 ) ]
\over
4 [ ( q p_1 ) ( q p_3 ) ]^{1/2} } \;\;\;.
\label{e11}
\end  {equation}
Here
$ \;
\displaystyle
Z = { 1 \over 2 } ( 1 \mp {\gamma}_5 )
\hat{q} = {\cal P}_{\mp}  \;\;  ,  \;\;\;  q^2=0.
$
    Massless 4-vector  $q$ can  be arbitrary,  but it  must be the
same for all considered nonclosed fermion lines of diagrams.

\begin {equation}
\displaystyle
\bar{u}_{\pm} (p_3) Q u_{\mp} (p_1)
= { Tr [ Q \hat{p}_1
( m \pm \hat{n} \hat{p} ) \hat{p}_3 ( 1 \mp {\gamma}_5 ) ] \over
4 \{ [ (p p_1) \pm m (np_1) ] [ (p p_3) \mp m (np_3) ] \} ^{1/2} }
\;\;\; .
\label{e12}
\end {equation}
Here
$ \;
\displaystyle
Z = { 1 \over  4m } ( m + \hat{p} ) ( 1 + {\gamma}_5 \hat{n} )
= {\cal P}, \;\;\;
p^2 =  m^2, \;\; n^2 = -1,  \;\; pn = 0.
$

    As regards 4-vectors $p$ and $n$ the same observation as the
one for vector $q$  in (\ref{e11}) is right.   We may require  for
maximum simplicity of calculations $ m = 0, \;\; p^2 = 0 \;$ in
(\ref{e12}).

    In  the   last  case   we  can   not  use   easier  operator
${\cal P}_{\pm}$ for $Z$, since  in  this  case  numerator and
denominator are identical with 0.

    If under numerical calculations denominator in (\ref{e11})  or
in (\ref{e12}) is equal to 0 for some values $p_1$ and  $p_3$,
it is sufficient to change values of arbitrary 4-vectors $q$  or
$p,n$ being  contained by  these formulae  (simultaneously for all
lines of diagrams being considered).

Another approach may be  useful  under  calculation  of  value
$\bar{u}_{\pm} (p_3) Q u_{\mp} (p_1)$:
\begin {equation}
\displaystyle
\bar{u}_{\pm} (p_3) Q u_{\mp} (p_1) \simeq
{ Tr [ Q {\hat p}_1 {\hat p}_3 ( 1 \mp {\gamma}_5 ) ]
\over 2 [ 2 ( p_1 p_3 ) ]^{1/2} } \;\; .
\label{e13}
\end {equation}
    Here $Z=1$ . However, it is  necessary to use formula (\ref{e6})  if
we have interfering amplitudes.  In this case [see (\ref{e5})]
\begin {equation}
\displaystyle
K = { Tr [ \hat{p}_1 \hat{p}_4 \hat{p}_2 \hat{p}_3
( 1 \mp {\gamma}_5 ) ] \over
8 [ (p_1 p_3) (p_1 p_4) (p_2 p_3) (p_2 p_4) ]^{1/2} } \;\; .
\label{e14}
\end {equation}
    Formulae (\ref{e13}), (\ref{e14}) generalize the method of
calculation of matrix elements  offered in \cite{r5}.

%REFERENCES

\end {document}